\documentclass[preprint,aps,12pt,notitlepage,nofootinbib,tightenlines]{revtex4}
\usepackage{amsmath}
\usepackage{enumerate, enumitem} 
\usepackage{bm}
\usepackage{braket}
\usepackage{color}
\usepackage{slashed}
\usepackage{multirow}
\usepackage{array}
\usepackage{float}
\usepackage{feynmp-auto}
\usepackage{booktabs}
\usepackage{threeparttable}
\usepackage{appendix}
\usepackage{ulem}

\usepackage{graphicx}

\definecolor{nicered}{rgb}{0.7,0.1,0.1}
\definecolor{nicegreen}{rgb}{0.1,0.5,0.1}
\usepackage[hidelinks]{hyperref}
\hypersetup{colorlinks=true,citecolor=nicegreen,linkcolor=nicered,urlcolor=magenta}

\newcommand{\be}{\begin{equation}\begin{aligned}}
\newcommand{\ee}{\end{aligned}\end{equation}}

\newcommand{\gev}{\text{ GeV}}
\newcommand{\tev}{\text{ TeV}}

\begin{document}

\title{Search for single vector-like $B$ quark production\\[0.05cm] in hadronic final states at the LHC}

\author{Bingfang Yang$^1$\footnote{Email: yangbingfang@htu.edu.cn}, Zejun Li$^1$\footnote{Email: lizejun@stu.htu.edu.cn}, Xinglong Jia$^1$\footnote{Email: jiaxinglong@stu.htu.edu.cn}, Stefano Moretti$^{2,3}$\footnote{Email: s.moretti@soton.ac.uk; stefano.moretti@physics.uu.se},
and Liangliang Shang$^{1,2}$\footnote{Email: shangliangliang@htu.edu.cn; liangliang.shang@physics.uu.se}}

\affiliation{\footnotesize
$^1$School of Physics, Henan Normal University, Xinxiang 453007, PR China\\
$^2$Department of Physics and Astronomy, Uppsala University, Box 516, SE-751 20 Uppsala, Sweden \\
$^3$School of Physics and Astronomy, University of Southampton, Highfield, Southampton SO17 1BJ, UK 
}

\vspace*{-1.25cm}

\begin{abstract}
\vspace*{0.5cm}
\baselineskip=15pt
{\small
In this paper, we study the discovery potential of a Vector-Like $B$ quark (VLB)  via the process $pp \to B(\to bZ)j\to b(Z \to \nu_l\bar{\nu_l})j$ at the Large Hadron Collider (LHC) with $\sqrt{s}=14$ TeV. In the framework of a simplified model, we perform a scan over its parameter space and test its viability following a Monte Carlo analysis developed to include all production and decay dynamics. We use cut-and-count combined with Extreme Gradient Boosting (XGBoost) methods to classify the signal and background events  in order to improve the efficiency of signal identification and background rejection. 
We find that this approach can reduce background events significantly while the signal retention rate is much higher than that of traditional methods, thereby improving the VLB discovery potential. We then calculate the exclusion and discovery capabilities for VLBs and find that 
the advantages of the cut-and-count plus XGBoost method especially lie in the high-mass region, i.e., $m_B > 1500 \gev$. We finally obtain the following LHC  results in terms of the coupling and chiral structure of a singlet heavy VLB  interactions: (i) for $g^{\ast}$=0.2 and $R_L=0$ with 3000 fb$^{-1}$,  the $B$ quark mass can be be excluded (discovered) up to 3000 GeV (2500 GeV); (ii) for $g^{\ast}$=0.2 and $R_L=0.5$ with 3000 fb$^{-1}$, the exclusion (discovery) region can reach up to 4750 GeV (4250 GeV).
}
\end{abstract}

\maketitle
\newpage

\section{Introduction}
The gauge hierarchy problem~\cite{Wilson:1970ag} caused by the quadratic divergence of the Higgs mass is one of the unsolved problems in the Standard Model (SM) of particle physics. For the past few decades, solving this problem has become a guiding principle for establishing New Physics (NP) models. Amongst these, Vector-Like Quarks (VLQs) are often introduced to alleviate this problem since an extra SM-like chiral fermion has been disfavored by the Higgs data collected at the Large Hadron Collider (LHC)~\cite{ATLAS:2019nkf,CMS:2018gwt}. 
VLQs are a type of hypothetical fermions with spin 1/2 and color-triplets, which  left- and right-handed components transform with the same properties under the SM Electro-Weak (EW) symmetry group~\cite{DeSimone:2012fs}.
There are seven possible multiplets of VLQs, depending on how they 
couple to the SM quarks~\cite{Aguilar-Saavedra:2013qpa}. These are EW singlets $[(T),(B)]$,  
doublets $[(X, T),(T, B),(B, Y)$] and triplets $[(X, T, B),(T, B, Y)]$, where the 
fields $X, Y, T, B$ have charges $+5/3e$, $-4/3e$, $+2/3e$ and $-1/3e$, respectively. Moreover,  VLQs can explain other anomalies observed in experiments~\cite{Branco:2022gja}, such as the $W$ mass anomaly~\cite{He:2022zjz,Cao:2022mif}. Many phenomenological analyses on VLQs have also been carried out~\cite{Vignaroli:2012sf, Vignaroli:2012nf, Yang:2019uea,Liu:2015kmo,Nutter:2012an,Deandrea:2021vje,Buckley:2020wzk,Gong:2019zws,Wang:2020ips,Yang:2022wfa,Han:2022rxy,Han:2021kcr,Shang:2022tkr,Cacciapaglia:2018qep,Carvalho:2018jkq,CarcamoHernandez:2021yev,Deandrea:2017rqp,Cacciapaglia:2018lld,Cacciapaglia:2017gzh,King:2020mau,Han:2024xyv,Banerjee:2023upj,Hernandez:2021tii,Bhardwaj:2022wfz, Bardhan:2022sif, Bhardwaj:2022nko}. 
While most such analyses where concerned with pair production of VLQs, as the limits on their masses grow stronger, single production of VLQs has also become of phenomenological interest \cite{Vignaroli:2012sf, Vignaroli:2012nf, Liu:2024hvp, Canbay:2023vmj,Han:2023jzm,Han:2022rxy,Han:2022jcp,Han:2022exz,Yang:2022wfa,Han:2022iqh,Han:2021lpg,Tian:2021nmj,Yang:2021btv,Deandrea:2021vje,Cetinkaya:2020yjf,Zhou:2020byj,Cacciapaglia:2018qep,Carvalho:2018jkq,Yang:2019uea,Liu:2016jho,Chen:2016yfv,Aguilar-Saavedra:2013qpa,Barcelo:2011wu,Shang:2024wwy,Liu:2019jgp}.
In particular, in Ref.~\cite{Han:2022iqh}, the authors have studied  single production of Vector-Like $B$ quarks (VLBs)  via the $bZ$ channel with $Z\to l^{+}l^{-}$ at the LHC. Herein, we will complement that study as we will be dealing with the same production process via  the $bZ$ channel but with $Z\to \nu\bar{\nu}$, again at the LHC but also in the presence of the up-to-date constraint  $m_B > 1500\text{ GeV}$.

In experiment, the ATLAS and CMS collaborations have implemented many analyses in the search for VLQs at the LHC. For VLBs, which  this work focuses on, the CMS collaboration has performed a search via pair production in the fully hadronic final state using Run 2 data with 138 fb$^{-1}$ and set  lower
limits on the VLB mass equal to 1570 GeV for BR$(B \to bH)$ = 100\% and 1540 GeV for BR$(B \to bZ)$
= 100\%~\cite{CMS:2020ttz}, i.e., in terms of VLB Branching Ratios  (BRs). For the singlet case, BR$(B \to bH)=$BR$(B \to bZ)$= 25\% and BR$(B \to tW)=50\%$, the lower
limit on the VLB mass is set to about 1060 GeV. For the $(B,Y)$ doublet case BR$(B \to bH)=$BR$(B \to bZ)$= 50\% and BR$(B \to tW)=0\%$, the lower
limit on the VLB mass is set to 1500 GeV. As intimated, although VLBs can be pair produced through strong 
interactions, the single production process through EW interactions becomes more important when 
VLBs are heavy, because of phase space effects. However, we also notice that the single VLQ production is model dependent, unlike the double VLQ one. The ATLAS collaboration has
performed a search for VLB single production decaying to a $Z$
boson and a Higgs boson using Run 2 data with 139 fb$^{-1}$~\cite{ATLAS:2023qqf}. 
For an isospin singlet VLB, the search has excluded values of the $c_W$ 
coupling\footnote{Note that, following Ref.~\cite{ATLAS:2023qqf}, $c_W$, $c_Z$, $c_H$ represents the coupling constants for interactions between the VLB and $W$, $Z$ and $H$ bosons, respectively. For singlet $B$ and $m_B > 1 \tev$, $c_W \approx \kappa$, $c_Z \approx \kappa/(\sqrt{2}\cos\theta_W)$, where $\theta_W$ is the Weinberg angle, and $c_H = m_B \kappa/(\sqrt{2}v)$, where $v$ is the Higgs vacuum expectation value, defined as $v \equiv 2m_W/g_W$, with $g_W$ representing the $SU(2)_L$ weak coupling. Here, $\kappa$ corresponds to $\kappa_B$ in Ref.~\cite{Buchkremer:2013bha}, where $\zeta_i=1$ and $c_W$ and $c_H$ do not contain the factor $g_W/2$. Finally,  in our paper, $\kappa$ corresponds to $g^*$ when $R_L=0$.}
greater than 0.45 for $m_B$ between 1.0 and 1.2 TeV. For $1.2 \text{ TeV} < m_B < 2.0 \tev$, $c_W$ values in the range $0.5 - 0.65$ have been excluded. For a VLB as part of a $(B,Y)$ doublet, the smallest excluded $c_Z$ coupling values range between 0.3 and 0.5 for $1.2 \text{ TeV} < m_B < 2.0 \tev$.

The NP processes can be rare and only differ slightly from the SM yield, a fact which represents a great challenge for physicists when analyzing data. Furthermore, data collected by LHC experiments are complex and highly 
dimensional whereas traditional data analysis techniques use a sequence of boolean decisions followed by statistical analysis on the selected data, both of which are based on the distribution of a single observed quantity motivated by physics considerations, which is not easily extended to the required higher dimensions~\cite{Guest:2018yhq}. As a result, physicists have done a lot of work to explore the application of Machine Learning (ML), including Deep Learning (DL), in particle physics and have developed a series of useful  tools~\cite{Kasieczka:2019dbj,Kagan:2020yrm,Chiang:2022lsn}. 
From these studies, we can see that ML has a natural advantage in identifying signals, wherein the Extreme Gradient Boosting (XGBoost)~\cite{Chen_2016} algorithm, stemming from Boosted Decision Trees (BDTs), has a unique advantage in finding NP effects~\cite{Alvestad:2021sje}. For these reasons, we will adopt a  XGBoost algorithm to improve the discovery potential of VLBs at the LHC, targeting not only Run 3 datasets, i.e., with 300 fb$^{-1}$ of luminosity, but also those of a
future High-Luminosity LHC (HL-LHC) \cite{Gianotti:2002xx}, i.e., with 3000 fb$^{-1}$ of luminosity.
 
We organize our paper as follows. In Sec. 2, we briefly describe the effective Lagrangian of the singlet VLB. In Sec. 3, we present the event generation and analysis of signal and backgrounds. In Sec. 4, we explore the exclusion and discovery potential at the (HL-) LHC. Finally, we summarize our results in Sec. 5.

\section{The effective Lagrangian}

In a simplified model including an $SU(2)$ singlet VLB, the Lagrangian that describes how the VLB couples to SM quarks and EW bosons can be expressed as follows~\cite{Buchkremer:2013bha}\footnote{In Ref.~\cite{Buchkremer:2013bha}, the effective Lagrangian for a singlet VLB is not clearly specified, yet, it can be obtained by incorporating the coefficients for the singlet VLB from Eq.~(3.2) in Ref.~\cite{Buchkremer:2013bha}. This Lagrangian is presented in the singlet VLB model of FeynRules~\cite{Alloul:2013bka}. However, there is a missing factor of $\sqrt{2}$ in the Lagrangian and the charge of $W$ is also inaccurately described on the website~\cite{feynrules}. Nevertheless, the provided codes are accurate, and we employed these in our work.}:
\begin{equation}
    \begin{split}
        \mathcal{L}_B  =& \frac{gg^*}{\sqrt{2}}\left\{\sqrt{\frac{R_L}{1+R_L}}\frac{1}{\sqrt{2}}\left[\bar{B}_L
        W_{\mu}^{-}\gamma^{\mu}u_L\right]+\sqrt{\frac{1}{1+R_L}}\frac{1}{\sqrt{2}}\left[\bar{B}_LW_{\mu}^{-}\gamma^{\mu}t_L\right]\right.\\
        & \left. + \sqrt{\frac{R_L}{1+R_L}}\frac{1}{2\cos\theta_W}\left[\bar{B}_L Z_\mu \gamma^\mu d_L\right]+\sqrt{\frac{1}{1+R_L}}
        \frac{1}{2\cos\theta_W}\left[\bar{B}_LZ_\mu \gamma^\mu b_L\right]\right.\\
        & \left. -\sqrt{\frac{R_L}{1+R_L}}\frac{m_B}{2m_W}\left[\bar{B}_RHd_L\right]-\sqrt{\frac{1}{1+R_L}}\frac{m_B}{2m_W}
        \left[\bar{B}_RHb_L\right]\right\}+h.c.
    \end{split}   
\end{equation}
Here, $g$ is the $SU(2)_L$ gauge coupling constant, $g^{\ast}$ parameterizes the single VLB production in association with 
SM quarks whereaa $m_W$ is the mass of the $W$ boson. Further, $R_L$ is the rate of the VLB decaying into the first and the third generation quarks. The decay widths of the VLB into SM quarks can be estimated as:
\begin{equation}
\begin{aligned}
\Gamma(B \to Zd_i) & =  \frac{g^2g^{*2}c^2_{d_i}}{512m^2_Zm^3_B\cos^2\theta_W}\left(
m^4_B + m^4_{d_i}+m^2_B m^2_Z-2m^4_Z+m^2_{d_i}(-2m^2_B+m^2_Z)\right) \\
&  \times \sqrt{m^4_{d_i}+\left(m^2_B-m^2_Z\right)^2-2m^2_{d_i}(m^2_B+m^2_Z)}, \\
\Gamma(B \to W^- u_i) & =  \frac{g^2g^{*2}c^2_{u_i}}{256m^2_Wm^3_B}\left( 
m^4_B + m^4_{u_i} + m^2_{u_i}m^2_W - 2m^4_W + m^2_B(-2 m^2_{u_i} + m^2_W) \right) \\
&  \times \sqrt{m^4_B+\left(m^2_{u_i} - m^2_W \right)^2 - 2m^2_B(m^2_{u_i} + m^2_W)}, \\
\Gamma(B \to Hd_i) & =  \frac{g^2g^{*2}c^2_{d_i}}{512m_W^2m_B}\left(
m^2_B + m^2_{d_i} - m^2_H \right)  \sqrt{m^4_{d_i}+\left(m^2_B - m^2_H \right)^2 - 2m^2_{d_i}\left( m^2_B + m^2_H \right).}
\end{aligned}
\end{equation}
Here, $i=1, 3$ denotes the generation of SM quarks, $c^2_{d_i,u_i}=R_L/(1+R_L)$ for $i=1$, $c^2_{d_i,u_i}=1/(1+R_L)$ for $i=3$. $m_{u_i}$ and $m_{d_i}$ are masses of the up-type and down-type SM quarks, respectively. Furthermore, $m_H$ is the mass of the SM-like Higgs boson.
If $R_L=0$, the VLB is assumed to couple only to the third generation of SM quarks~\cite{Aguilar-Saavedra:2013qpa} and the Lagrangian can be simplified as:
\begin{equation}
    \begin{split}
    \mathcal{L}_B  =& \frac{gg^*}{\sqrt{2}}\left\{\frac{1}{\sqrt{2}}\left[\bar{B}W_\mu^{-}
    \gamma^{\mu}t_L\right]+\frac{1}{2\cos\theta_W}\left[\bar{B}Z_\mu^{-}\gamma^{\mu}
    b_L\right]\right.\\
    & \left. -\frac{m_B}{2m_W}\left[\bar{B}_RHb_L\right]-\frac{m_B}{2m_W}\left[\bar{B}_L
    Hb_R\right]\right\}+h.c.
    \end{split}  
\end{equation}
For $m_B>1\text{ TeV}$, the decay BR of the VLB tends
to be such that BR$(B \to tW)$:BR$(B \to bZ)$:BR$(B \to bH)$ = 2:1:1, as expected
from the Goldstone equivalence theorem~\cite{He:1992nga,He:1993yd}.

\begin{figure}[t!]
		\includegraphics[scale=0.85]{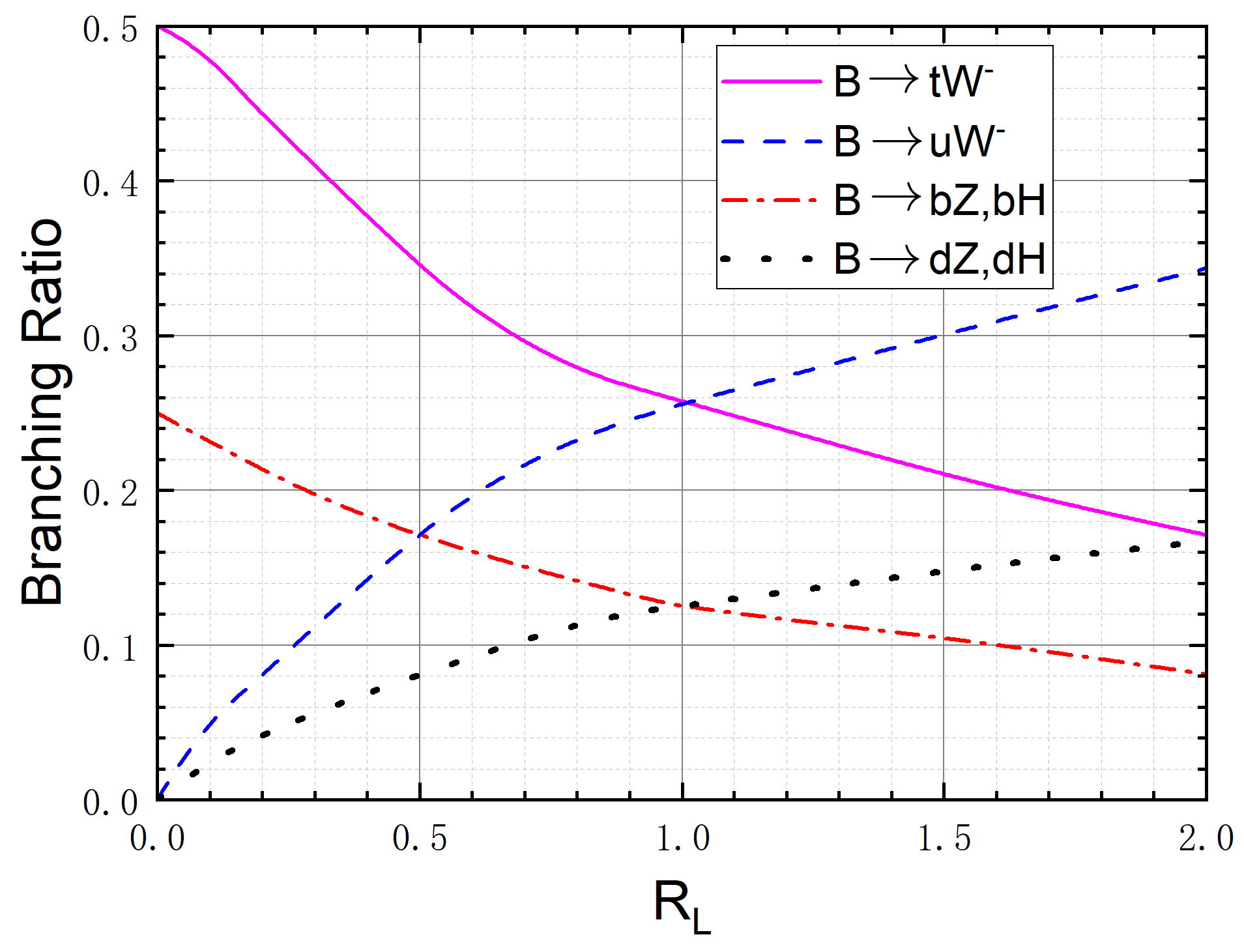}
		\includegraphics[scale=0.85]{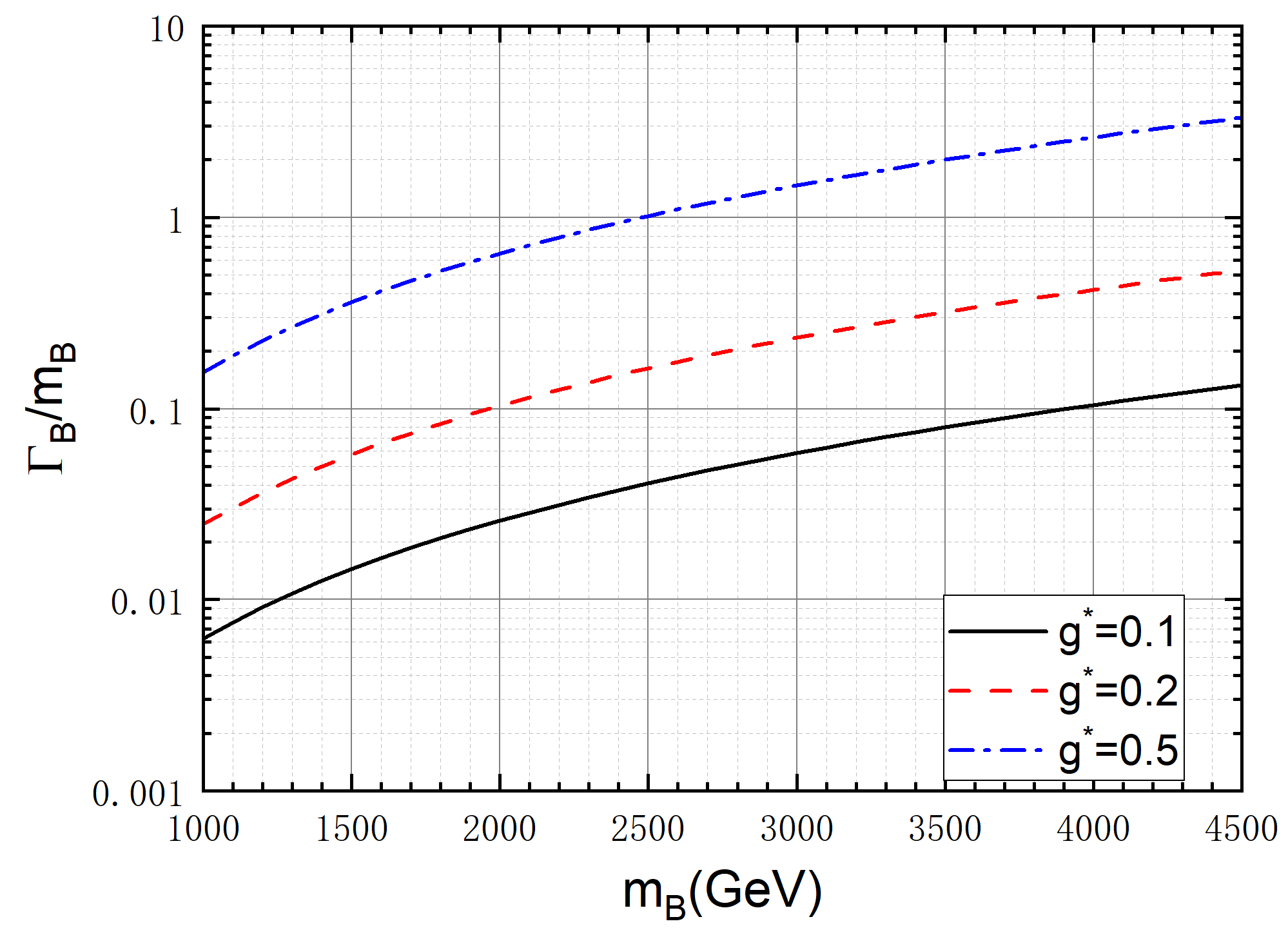}
    \caption{The BRs of the VLB as a function of $R_L$ with $m_{B}$=1500 GeV (left) and the width-to-mass ratios of the VLB as a function of $m_{B}$ (right).}
    \label{branching ratios}
\end{figure}

In Fig.~\ref{branching ratios}, we show the BRs of our VLB as a function of $R_L$ (left) for $m_{B}$=1500 GeV and the width-to-mass ratios as a function of $m_{B}$ (right). The BRs are independent of the coupling coefficient $g^{\ast}$ whereas the width-to-mass ratios are approximately independent of $R_L$ when $m_B > 1 \tev$. For a fixed $g^{\ast}$, the width-to-mass ratios are proportional to $m_{B}^2$. We can see that the VLB mainly decays to the third generation quarks for $R_L < 0.5$. However, when $R_L > 1$, the VLB decay to the first generation quarks gradually becomes the dominant decay mode. In this work, we primarily focus on the VLB coupling to the third generation quarks, hence, we consider two cases: $R_L=0$ and $R_L=0.5$. Since the constraints from the EW Precision Observables (EWPOs) on the coupling coefficient $g^{\ast}$ are of  $\mathcal{O}(10^{-1})$ \cite{Cao:2022mif}, we consider here the range $g^{\ast}\leq 0.5$.

\begin{figure}[h!]
    \centering
 \includegraphics[scale=0.33]{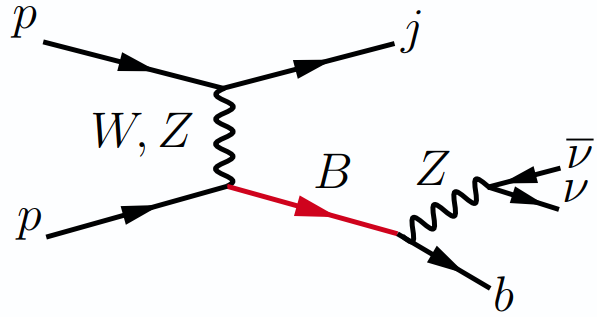}
		\includegraphics[scale=0.33]{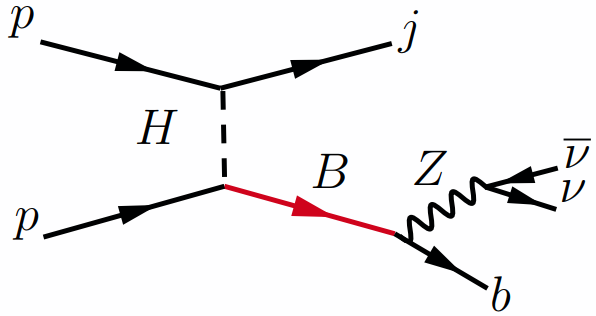}    
            \caption{Feynman diagrams for single VLB (in red) production with subsequent decay into $bZ$ at the LHC.}\label{feynman diagram}
\end{figure}

We explore the discovery potential of single VLB production through the channel
$pp \to B(\to bZ)j\to b(Z \to\nu_l\bar{\nu_l})j$,
where $b$ denotes the $b$-tagged jet and $j$ stands for light-flavor jets. We show the typical Leading Order (LO) Feynman diagrams in Fig.~\ref{feynman diagram}. From this figure, we can see that the signal events are characterized by two jets with at least one $b$-tagged jet among these, along with significant missing (transverse) energy. Therefore, the primary SM backgrounds are $pp \to Zjjj$, $pp\to bZj$, $pp\to ZZ$ and $pp\to ZH$.
Among these, $pp\to bZj$ is an irreducible background while the others are considered reducible backgrounds. The latter must be taken into account due to jets exhibiting lower tagging efficiencies and higher mistagging rates compared to leptons at hadron colliders.
We have also evaluated further backgrounds including $pp \to W^\pm Z$ and determined that their impact can be disregarded according to the selection criteria that will be discussed later.

\begin{figure}[h!]
		\includegraphics[scale=0.85]{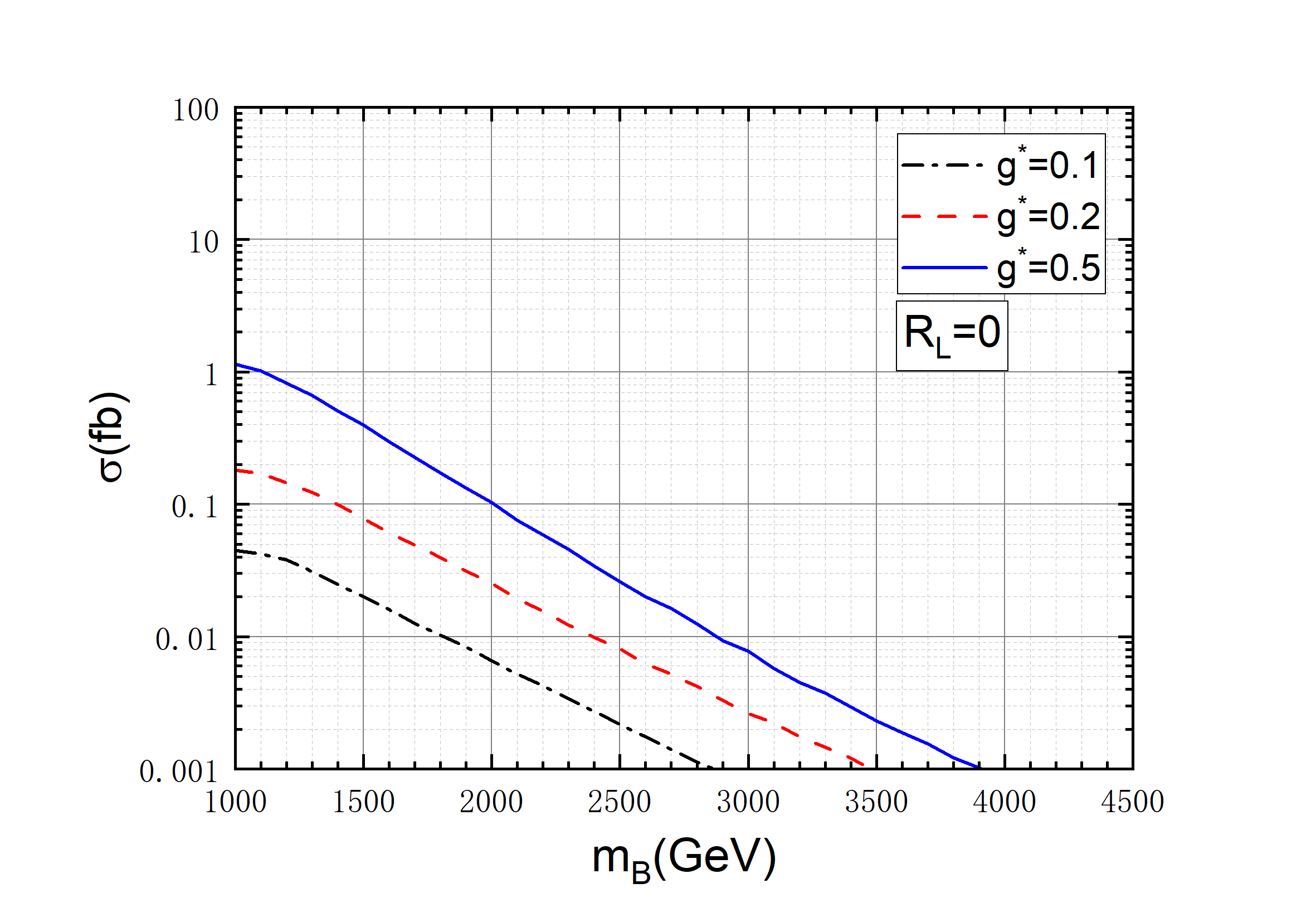}
		\includegraphics[scale=0.85]{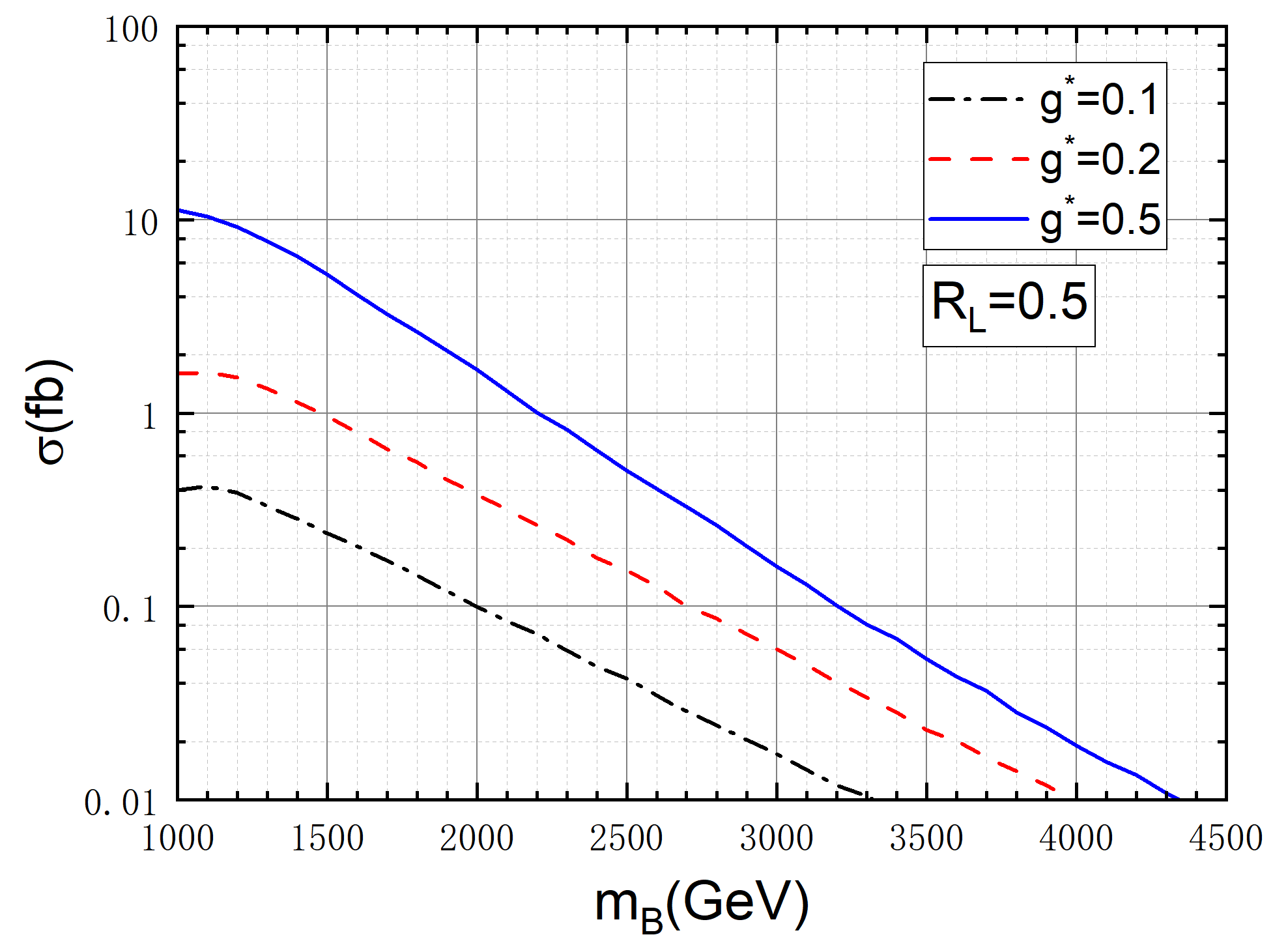}
	\caption{Cross sections of the signal process $pp \to B(\to bZ)j\to b(Z \to\nu_l\bar{\nu_l})j$ as a 
		function of $m_B$ at the 14 TeV LHC with $g^{\ast}$ = 0.1, 0.2, 0.5 for $R_L$=0 (left) and $R_L$=0.5 (right). Here, the conjugate processes have been included. }\label{Cross sections}
\end{figure}    

We calculate the LO cross sections of the signal process by MadGraph5\_aMC@NLO \cite{Alwall:2014hca} with the NNPDF23NLO 
Parton Distribution Functions (PDFs) \cite{NNPDF} and  default renormalization and factorization scales.  The cross sections at 14 TeV LHC are depicted as a function of $m_B$ in Fig.~\ref{Cross sections}, where we take $g^{\ast}$ = 0.1, 0.2 and 0.5 as well as $R_L=0$ and $0.5$ as examples. The input SM parameters are taken as follows\footnote{Here, $m_W$ is calculated as $\sqrt{\frac{M_Z^2}{2} + \sqrt{\frac{M_Z^4}{4} - \frac{\alpha_{\text{EW}}\pi M_Z^2}{G_F\sqrt{2}}}}$ within MadGraph5\_aMC@NLO.}~\cite{ParticleDataGroup:2022pth}:
\begin{center}
$m_b=4.18 \gev$, $m_Z=91.1876 \gev$, $m_H = 125.25 \gev$,\\
$G_F = 1.166378 \times 10^{-5} \text{ GeV}^{-2}$, $\sin^{2}\theta_W=0.231$, $\alpha_{\text{EM}}(m_Z)=1/128$.
\end{center}
From Fig.~\ref{Cross sections}, we can see that the signal cross sections decrease rapidly with increasing $m_B$. For a fixed $m_B$, the cross sections increase rapidly with increasing $g^{\ast}$ due to $\sigma \propto g^{\ast2}$. 
Furthermore, for equivalent values of the coupling $g^*$ and the mass $m_B$, the cross section for $R_L = 0.5$ is larger than that for $R_L = 0$. This is because the former scenario involves $B$ mixing with the first generation quarks, which significantly enlarges the single VLB production rate due to the increased sea quark PDFs. However, we limit ourselves to studying cases where $R_L < 0.5$ because there are stringent constraints from non-LHC flavor physics data on VLQs mixing significantly with lighter quarks \cite{Belfatto:2023tbv, Crivellin:2022rhw, Balaji:2021lpr, Branco:2021vhs, Vatsyayan:2020jan, Botella:2016ibj, Ishiwata:2015cga, Alok:2015iha, Cacciapaglia:2015ixa, Botella:2012ju, Cacciapaglia:2010vn}. We list the LO cross sections and the $K$ factors representing the QCD corrections for the background processes in Tab.~\ref{Kyinzi}.

\begin{table}[!t]
\centering
\begin{tabular}{ccccc}
\hline
Processes & $Zjjj$    & $bZj$  & $ZZ$ & $ZH$  \\ \hline
$\sigma_{\text{LO}}$(fb) & 22.5 & 5.8 & 0.1 & 0.1 \\ \hline
$K$ factor  & 1.2~\cite{Alwall:2014hca}   & 1.2~\cite{Campbell:2005zv}  & 1.8~\cite{Campbell:1999ah} & 1.3~\cite{Alwall:2014hca}  \\ \hline
\end{tabular}
\caption{The LO cross sections and the $K$ factors representing the QCD corrections for the background processes at the 14 TeV LHC. Here, the cross section of $bZj$ includes the contribution from the conjugate process of $pp \to \bar{b}Zj$.}
\label{Kyinzi}
\end{table}

\section{Event generation and analysis}       
We conduct a detailed detector simulation in this work. The signal model file is obtained from \text{FeynRules} \cite{feynruls} and, again,  parton-level events are generated using MadGraph5\_aMC$@$NLO with the NNPDF23NLO PDF set. Indeed, the factorization and renormalization scales are still set to their default values in MadGraph5\_aMC$@$NLO.

These events are then input into Pythia 8.3~\cite{Bierlich:2022pfr} for parton showering and hadronization. Subsequently, fast detector simulations are performed using \text{Delphes 3.4.2} \cite{deFavereau:2013fsa} with the built-in detector configurations of the LHC Run 3 and HL-LHC~\cite{website_hllhc}. Jets are clustered using \text{FastJet} \cite{Cacciari:2011ma} employing the anti-$kt$ algorithm \cite{Cacciari:2005hq}. Settings for the cone radius of the jet tagging algorithm and isolated leptons are implemented in Delphes with default values. Then, both signal and background events are analyzed using \text{MadAnalysis 5} \cite{Conte:2012fm}. Finally, we scan the VLB parameter space and connect these programs with the help of the EasyScan\_HEP package \cite{Shang:2023gfy}. To quantify the observability of the signal, we utilize the following formula to calculate the expected discovery and exclusion significance, without considering systematic uncertainties for simplicity \cite{Burns:2011xf}:
\begin{equation}
    \mathcal{Z}_{\rm disc}=\sqrt{2\left[\left(s+b\right)\ln\left(1+\frac{s}{b}\right)-s\right]},
\end{equation}
and
\begin{equation}
    \mathcal{Z}_{\rm excl}=\sqrt{2\left[s-b\ln\left(1+\frac{s}{b}\right)\right]},
\end{equation}
where $s$ and $b$ are the number of signal ($s$) and total background ($b$) events, respectively, after applying the selection criteria that will be discussed later. The exclusion significance is denoted by $\mathcal{Z}_{\rm excl}$ = 2 whereas the discovery significance is denoted by $\mathcal{Z}_{\rm disc}$ = 5.

To accommodate contributions from higher-order QCD corrections, we adjust the cross sections of the dominant backgrounds from the LO to the Next-to-LO (NLO) or Next-to-NLO (NNLO) using $K$ factors, as detailed in Tab.~\ref{Kyinzi}. We assume that kinematic distributions are minimally impacted by these higher-order effects. Therefore, for simplicity, we scale the kinematic distributions, which we will discuss later, using constant bin-independent $K$ factors. It is then worth noting that, for the signal, we maintain the LO cross section, ensuring that the exclusion and discovery potentials depicted in this analysis remain conservative. The inclusion of higher-order QCD corrections could indeed bolster these potentials.

For our analysis, we establish as reference values for the couplings $g^* = 0.2$ and $R_L = 0$. Furthermore, along the VLB mass direction, our Benchmark Points (BPs) are located at the $m_B$ values of 1500 GeV and 2000 GeV. However, we will later showcase the excluded $2\sigma$ and $5\sigma$ discovery thresholds in the $g^* - m_B$ plane, considering two different values of $R_L$, namely 0 and 0.5 and more finely sifted $m_B$ values. 

\begin{figure}[t!]
        \includegraphics[width=0.45\linewidth]{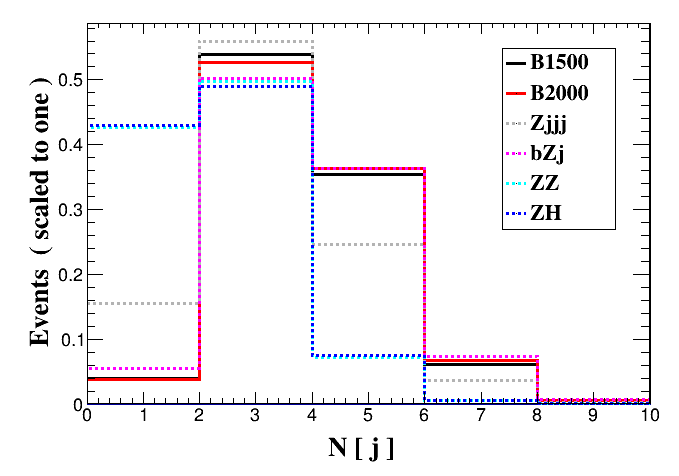}
        \includegraphics[width=0.45\linewidth]{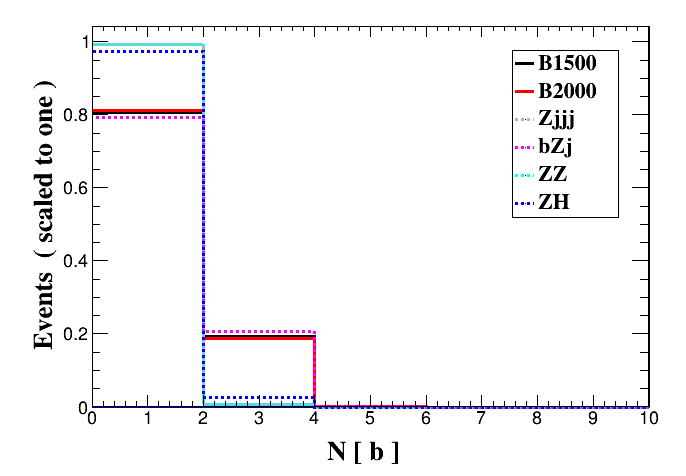}
        \includegraphics[width=0.45\linewidth]{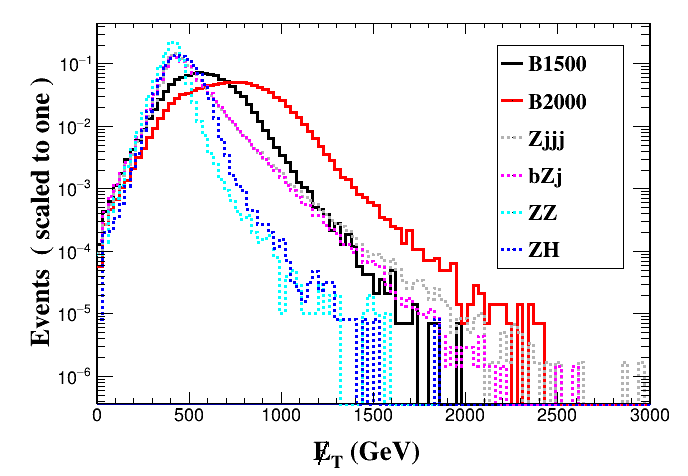}
        \includegraphics[width=0.45\linewidth]{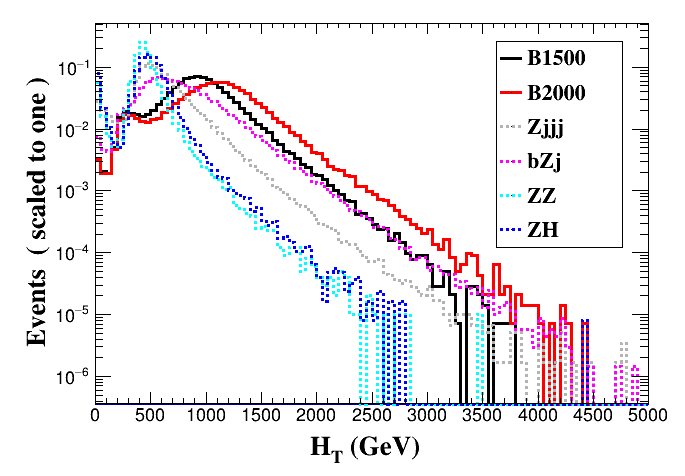}
        \includegraphics[width=0.45\linewidth]{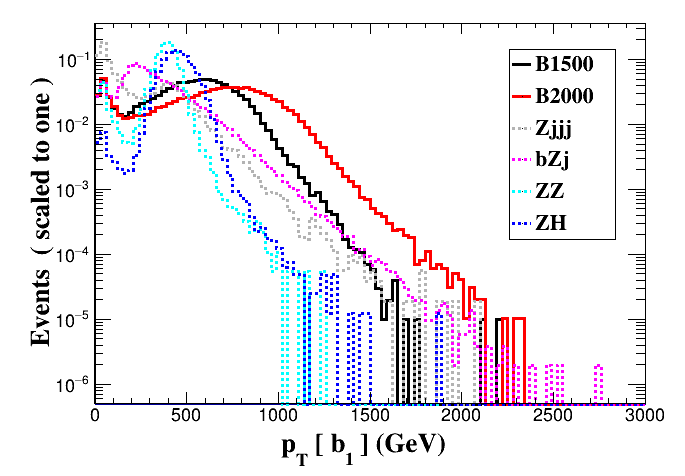}
        \includegraphics[width=0.45\linewidth]{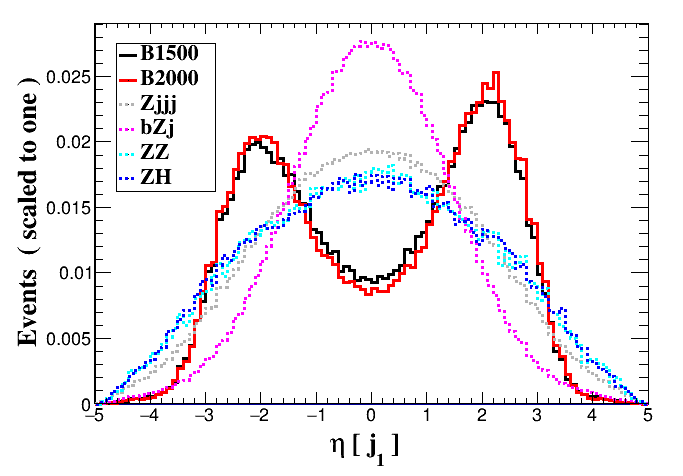}
        \includegraphics[width=0.45\linewidth]{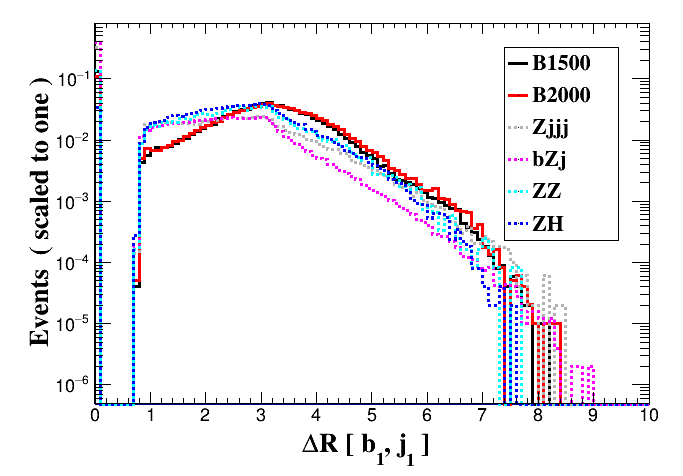}
    \caption{Normalized distributions for the signals ($m_B=1500$ GeV and 2000 GeV are selected as examples) and backgrounds at the 14 TeV LHC for $g^{\ast}$ = 0.2 and $R_L=0$.}\label{distributions}
\end{figure}

Due to the large mass of the VLB, its decay products are highly boosted, this  resulting in the transverse momentum ($p_T$) peaks of the signals being larger than those of the (SM) backgrounds. In Fig.~\ref{distributions}, we present several differential distributions for signals and backgrounds at the LHC, including the transverse momentum spectrum of the leading $b$-tagged jet ($p_T^{b_1}$), the distribution of total transverse hadronic energy $H_T$ (defined as $H_T \equiv \sum_{\text{hadronic particles}}|\vec{p}_T|$) and the that of the missing transverse energy $\slashed{E}_T$. Additionally, the signal process involves an outgoing light quark, leading to a light-flavor forward jet within the detector. Therefore, we also display kinematic distributions showcasing distinct spatial characteristics, such as the pseudorapidity distribution of the leading light-flavor jet $\eta_{j_1}$ and the distribution of spatial separation $\Delta R$ between the leading $b$-tagged jet and the leading light-flavor jet. The spatial separation is defined as $\Delta R = \sqrt{\Delta \Phi^{2}+ \Delta  \eta ^{2} }$, representing the separation in the rapidity ($\eta$)–azimuth ($\phi$) plane.

Then, we employ the XGBoost algorithm to enhance the discovery potential of the VLB at the LHC. We select four signal BPs with $m_B$ values of 1500 GeV, 2000 GeV, 3000 GeV and 4000 GeV while fixing $g^*=0.1$ and $R_L = 0$ throughout, as kinematic distributions are minimally affected by the coupling $g^*$ and the mixing parameter $R_L$. The background events are generated for the following processes here: 
\begin{itemize}
    \item  $pp \to Z(\to \nu_l\bar{\nu_l})jjj$,
    \item  $pp\to bZ(\to \nu_l\bar{\nu_l})j$,
    \item  $pp\to Z(\to \nu_l\bar{\nu_l})Z(\to b\bar{b}/jj)$,
    \item  $pp\to Z(\to \nu_l\bar{\nu_l})H(\to b\bar{b})$.
\end{itemize}
Considering the characteristics of the signal distributions, for the purpose of optimising the Monte Carlo (MC) generation efficiency, 
events  in this analysis must satisfy $p_T^{j,b} > 200 \gev$ and $\slashed{E}_T > 400 \gev$ at the parton level\footnote{We have indeed checked that these are not biasing the detector level events, see Cut-2 and -4 below.}, implemented using MadGraph5\_aMC$@$NLO, and $N(j) \geq 1$ and $N(b) \geq 1$ at the detector level, implemented using MadAnalysis 5.

Each event from the signal BPs and the four types of backgrounds corresponds to a 10-dimensional event vector constructed by 9-dimensional input features and a 1-dimensional target variable. The input features consist of the number of $b$-tagged jets $N(b)$, the number of light-flavor jets $N(j)$, the transverse momentum ($p_T^{b_1}$) and pseudorapidity ($\eta_{b_1}$) of the leading $b$-tagged jet, the transverse momentum ($p_T^{j_1}$) and pseudorapidity ($\eta_{j_1}$) of the leading light-flavor jet, the spatial separation between the leading $b$-jet and light-jet $\Delta R(b_1, j_1)$, the total transverse hadronic energy $H_T$ and the missing energy $\slashed{E}_{T}$, obtained with the assistance of expert mode in MadAnalysis 5. The target variable is set as 1 for signal events and 0 for background events.
We employ $8 \times 10^5$ event vectors, with equal contributions from each signal BP and background process, as  training data. This dataset enables the XGBoost model (hereafter referred to as the X-model) to learn the relationships between features and how to map these onto target variables. Additionally, we use $1.6 \times 10^4$ event vectors, with equal contributions from each signal BP and background process, as testing data to evaluate the performance of the trained X-model on new signal and background events different from those in the training data.

There are hyperparameters used to fine-tune the performance of the X-model. The learning rate $\alpha$ controls the step size during the minimization of a logistic loss function $\mathcal{F}$:
\begin{equation}
\mathcal{F} = -\frac{1}{N} \sum_{i=1}^{N} \left[ y_i \ln(p_i) + (1 - y_i) \ln(1 - p_i) \right].
\end{equation}
Here, $N$ represents the number of events used to train the X-model, $y_i$ denotes the target variable for each event and $p_i$ indicates the predicted probability of the signal class for the $i$-th event. 
The fractions of event vectors ($r_\text{s}$) and features ($r_\text{f}$) in each event vector can be randomly selected for each tree. This randomness is introduced into the X-model to enhance its generality.
Additionally, the parameter $\gamma$ prevents overfitting by penalizing excessively large and deep trees. If $\mathcal{F}>\gamma$, a further partition on a leaf node of the tree is required. 
We specify the learning rate to range between 0.01 and 0.5. The ranges for the fractions of event vectors and features are established as $r_\text{s,f} \in [0.5, 1.0]$. Furthermore, the ranges for the number of trees $n_{\text{t}}$ and the maximum depth of each tree $n_{\text{d}}$ are set as [50, 201] and [3, 21], respectively. A summary of these hyperparameter ranges and the optimal set of hyperparameters is provided in Tab.~\ref{hyperp}.

\begin{table}[!t]
\centering
\begin{tabular}{ccccccc}
\hline
Hyperparameter & $\alpha$ & $\gamma$  & $n_{\text{t}}$ & $n_{\text{d}}$ & $r_\text{s}$ &  $r_\text{f}$ \\ \hline
Range & [0.01, 0.3]  & [1, 5]  & [20, 201] & [3, 21] & [0.5, 1] & [0.5, 1] \\ \hline
Optimal value & 0.2 & 1.3 & 183 & 15 & 0.96 & 0.99 \\ \hline
\end{tabular}
\caption{The hyperparameter ranges and optimal values considered in this work.}
\label{hyperp}
\end{table}

\begin{figure}[h!]
        \includegraphics[width=7.9cm , height=6.3cm]{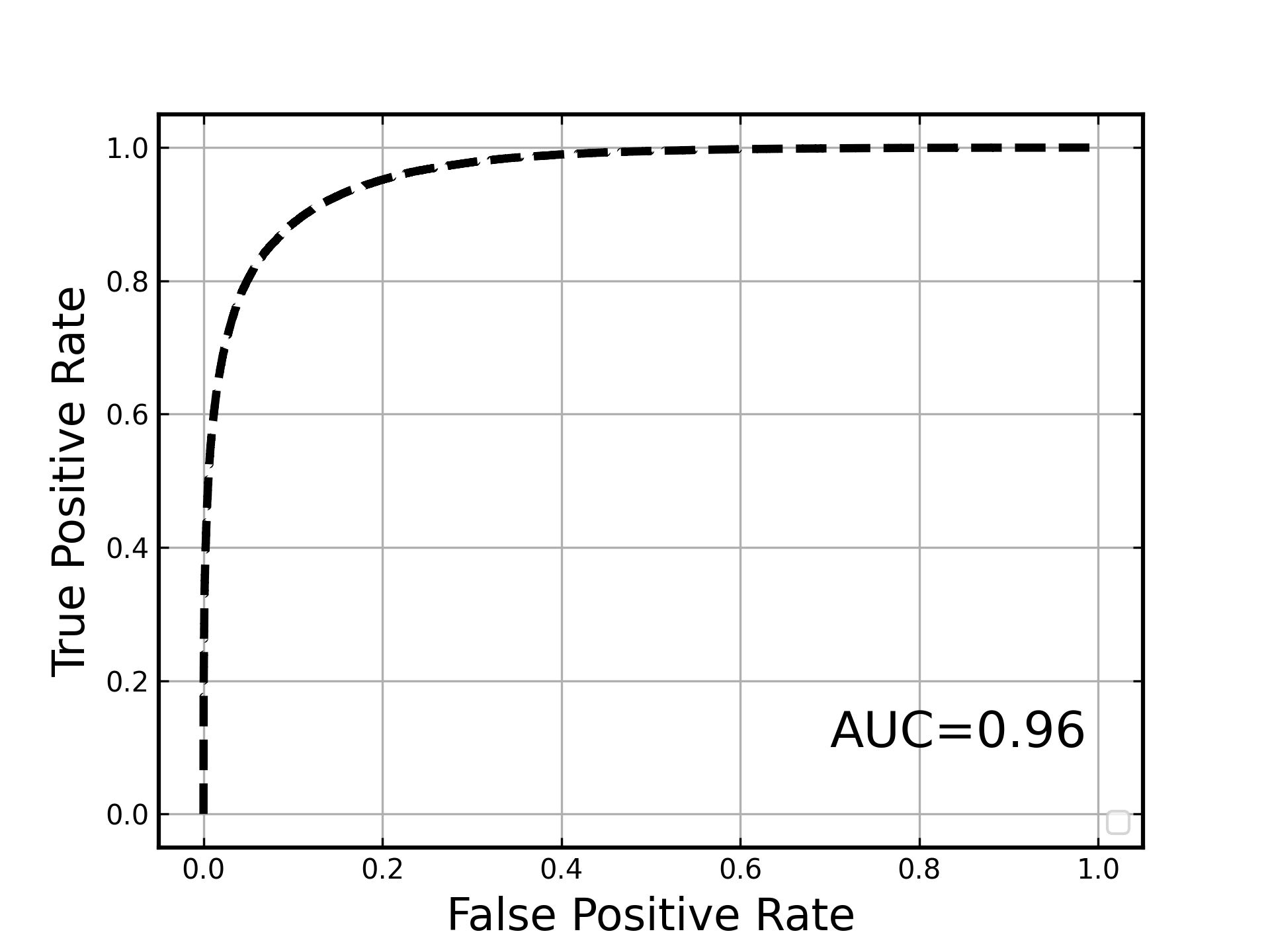}
        \includegraphics[width=7.9cm , height=6.3cm]{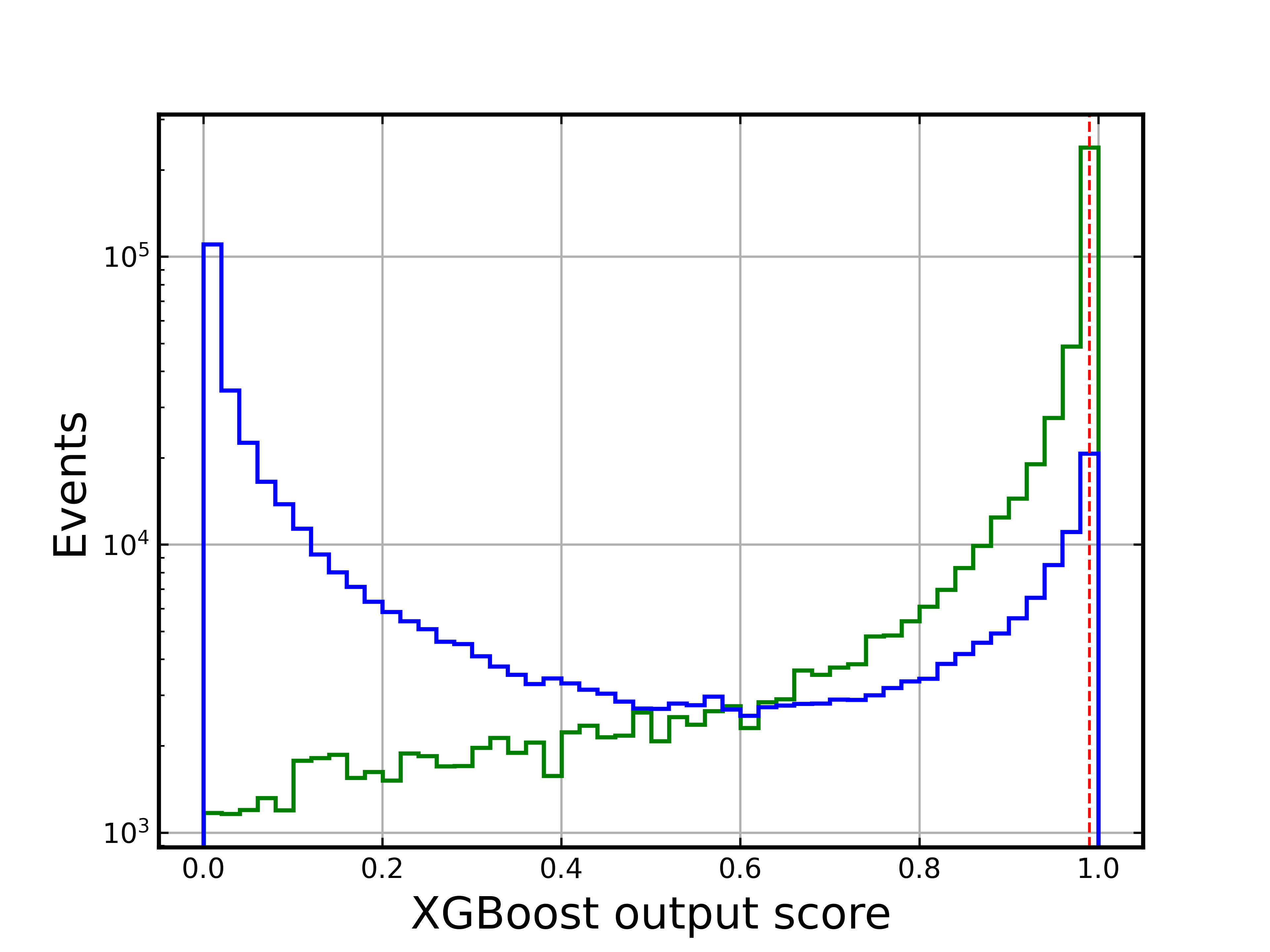}
    \caption{Model performance on testing data. The left is the ROC curve while AUC represents the area under the curve. The right is the score distributions relevant to testing data. The green line stands for the signal while the blue line stands for all the backgrounds. The vertical red line stands for the threshold we choose.}\label{model performance}
\end{figure}

When assessing the performance of a set of hyperparameters, each $p_i$ and $p_{i+1}$ serves as threshold probabilities denoted by $\mathcal{T}_j$ ($j=1,\dots,N+1$). For every $\mathcal{T}_j$, if $p_i > \mathcal{T}_j$ ($i=1,\dots,N$), the corresponding event is classified as a signal, otherwise, it is classified as background. Consequently, each set of hyperparameters is associated with a Receiver Operating Characteristic (ROC) curve, which illustrates the True Positive Rate (TPR) against the False Positive Rate (FPR) at various threshold settings. Here, $j$ ranges from 1 to $N+1$ and the Area Under the ROC Curve (AUC) \cite{FAWCETT2006861} determines the effectiveness of the hyperparameters.
Bayesian optimization, as introduced by  \cite{snoek2012practical}, is employed to discover the optimal set of hyperparameters for this study. The objective function chosen for this optimization is the AUC value.
During each iteration of Bayesian optimization, the training event vectors are randomly split into a 4:1 ratio. The former portion is utilized to train the X-model using XGBoost, while the latter is reserved for guiding the selection of the next set of hyperparameters to evaluate via Bayesian optimization.
Subsequently, the optimal set of hyperparameters is identified based on maximizing the AUC value. Once this optimal set is determined, it represents the desired configuration for the X-model, which we refer to as the trained X-model.
The trained X-model is then applied to the testing data as discussed earlier. Subsequently, the threshold probability closest to the point (1,0) on the resulting ROC curve is used as the selection criterion of the XGBoost algorithm to distinguish signal from background events used to achieve discovery or exclusion. We denote this threshold probability as $\mathcal{T}^*$.
Ultimately, we utilize the trained X-model to generate the ROC curve and probability distributions of X-model outputs for events in the testing data. These plots, along with the selection criterion $\mathcal{T}^*$, are depicted in Fig.~\ref{model performance}.

\section{LHC Exclusion and Discovery Potential}

\begin{table}[htb]
\centering
\begin{tabular}{p{1.6cm}<{\centering} p{2.0cm}<{\centering} p{2.0cm}<{\centering} p{0.3cm}<{\centering} p{2.2cm}<{\centering} p{2.2cm}<{\centering}  p{2.2cm}<{\centering} p{2.2cm}<{\centering}}
\toprule[1.5pt]
\multirow{2}{*}{Cuts}& \multicolumn{2}{c}{Signals}&\multicolumn{5}{c}{Backgrounds}  \\ \cline{2-3}  \cline{5-8}
 &1500 GeV &2000 GeV && $Zjjj$    & $bZj$  & $ZZ$ & $ZH$ \\  \cline{1-7} 
\midrule[1pt]
Initial & 0.079 & 0.024 && 26.94 & 6.26 & 0.14 & 0.11 \\
Cut-1 & 0.056 & 0.017 && 2.26 & 4.55 & 0.017 & 0.046 \\
Cut-2 & 0.030 & 0.011 && 0.24 & 0.77 & 0.0009 & 0.013 \\
Cut-3 & 0.015 & 0.006 && 0.07 & 0.13 & 0.0002 & 0.0046 \\
Cut-4 & 0.010 & 0.005 && 0.02 & 0.04 & $4.26\times 10^{-6}$ & 0.0001 \\ \hline
Efficiency & 12.9\% & 20.8\% && $7.4\times 10^{-4}$ & $6.8\times 10^{-3}$ & $3.0\times 10^{-5}$ & $9.0\times 10^{-4}$ \\
\bottomrule[1.5pt]
\end{tabular}
\caption{Cut flow of the cross sections (in fb) for the signals with two typical VLB quark masses and backgrounds at the 14 TeV LHC using the cut-and-count method. Here we take the parameters $g^{\ast}=0.2$ and $R_L=0$.}
\label{cutflow}
\end{table}

Drawing from the kinematic distributions illustrated in Fig.~\ref{distributions}, we implement the subsequent kinematic selection criteria on the events to distinguish the signal from the backgrounds:
\begin{enumerate}[label=\textbf{Cut-\arabic*:}, ref=\arabic*, start=1]
\item There are at least two jets ($N(j,b)\geq 2$) and at least one $b$-tagged jet among these ($N(b)\geq 1$).
\item The transverse momentum of the leading $b$-tagged jet is required to be greater than $500 \gev$ ($p_T^{b_1} > 500 \gev$).
\item The pseudorapidity of the leading light-flavor jet is required to be greater than $1.8$ ($|\eta_{j_1}| > 1.8$). Additionally, the spatial separation between the leading $b$-tagged jet and the leading light-flavor jet is required to be greater than $2.0$ to reduce background from the process $pp \to bZj$, which contains the $b$ quark in the initial state ($\Delta R(j_1, b_1) > 2.0$).
\item The total transverse hadronic energy is required to be greater than $800 \gev$ ($H_T > 800 \gev$) and the transverse missing energy is required to be greater than $500 \gev$ ($\slashed{E}_T > 500 \gev$).
\end{enumerate}

\begin{table}[htb]
\centering
\begin{tabular}{p{2.0cm}<{\centering} p{2.0cm}<{\centering} p{2.0cm}<{\centering} p{0.3cm}<{\centering} p{2.0cm}<{\centering} p{2.0cm}<{\centering}  p{2.0cm}<{\centering} p{2.0cm}<{\centering}}
\toprule[1.5pt]
\multirow{2}{*}{Cuts}& \multicolumn{2}{c}{Signals}&\multicolumn{5}{c}{Backgrounds}  \\ \cline{2-3}  \cline{5-8}
 &1500 GeV &2000 GeV && $Zjjj$    & $bZj$  & $ZZ$ & $ZH$ \\  \cline{1-7} 
\midrule[1pt]
Initial & 0.079 & 0.024 && 26.94 & 6.26 & 0.14 & 0.11 \\
Basic cuts & 0.056 & 0.017 && 2.26 & 4.55 & 0.017 & 0.046 \\
$\mathcal{T}^*$ & 0.006 & 0.005 && 0.0075 & 0.019 & 0.000005 & 0.00002 \\
Efficiency & 7.5\% & 20.8\% && $2.8\times 10^{-4}$ & $3.0\times 10^{-3}$ & $3.6\times 10^{-5}$ & $1.8\times 10^{-4}$ \\
\bottomrule[1.5pt]
\end{tabular}
\caption{Cut flow of the cross sections (in fb) for the signals with two typical VLB quark masses and backgrounds at the 14 TeV LHC using the XGBoost method. Here we take the parameters $g^{\ast}=0.2$ and $R_L=0$.}
\label{cutflow2}
\end{table}

We present the cross sections of two representative signals, with $m_B$ values set at 1500 GeV and 2000 GeV, along with the corresponding backgrounds. These are depicted after applying the selection criteria detailed in Tab.~\ref{cutflow} and Tab.~\ref{cutflow2} for the cut-and-count method and XGBoost method, respectively. To construct these tables, we have simulated $10^5$ events at the detector level for each signal and background process.

From Tab.~\ref{cutflow}, we can see that, among the four types of backgrounds,  the dominant ones are the $Zjjj$ and $bZj$ prior to any cuts. The first two cuts, pertaining to the number of final jets and transverse momentum of the leading $b$-tagged jets, significantly cut events from $Zjjj$ and $ZZ$, achieving a cumulative reduction rate of approximately 99\%. Subsequently, the third cut, addressing pseudorapidities and spatial separations, efficiently cuts $bZj$ events with a reduction rate of about 83\%. Furthermore, imposing stringent requirements on $H_T$ and $\slashed{E}_T$ successfully cuts overall around 70\% of background events while retaining 70\% of signal ones. Altogether, at the end of the cut flow, all backgrounds undergo notable suppression whereas the signals maintain a relatively favorable efficiency. In the end, the primary background contributions originate from the $Zjjj$ and $bZj$ SM processes, which remain comparable in magnitude to the signal remnants, exhibiting cross sections of $0.02 \text{ fb}$ and $0.04 \text{ fb}$, respectively. Finally, compared to the cut-and-count method, from Tab.~\ref{cutflow2}, we can see that the signal significance is clearly improved further by the XBoost method.

To determine the exclusion and discovery reach at the 14 TeV LHC, using the two methods adopted here (i.e., the traditional cut-and-count one and the ML one), we perform calculations spanning the interval $m_{B}=1500\text{ GeV}$ to 3000 GeV. BPs are now  selected at intervals of 250 GeV in the VLB mass, with $R_L=0$ and $g^*=0.2$. For each signal BP, we simulate $10^5$ events, while for each background, we simulate $10^6$ events. It has been verified that different values of $R_L$ and $g^*$ yield the same efficiency when $m_B$ remains constant.

\begin{figure}[h!]
    \includegraphics[scale=0.85]{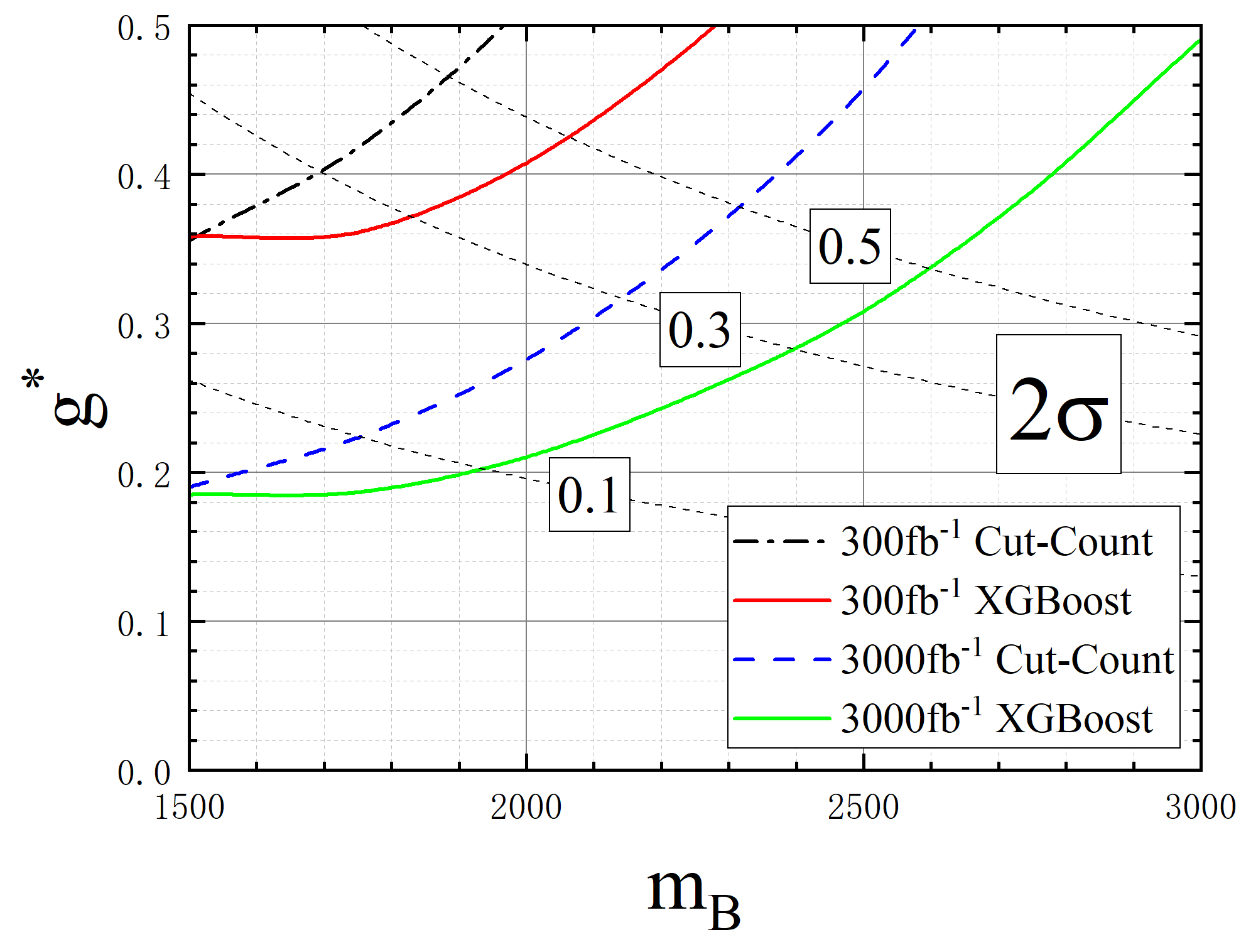}
    \includegraphics[scale=0.85]{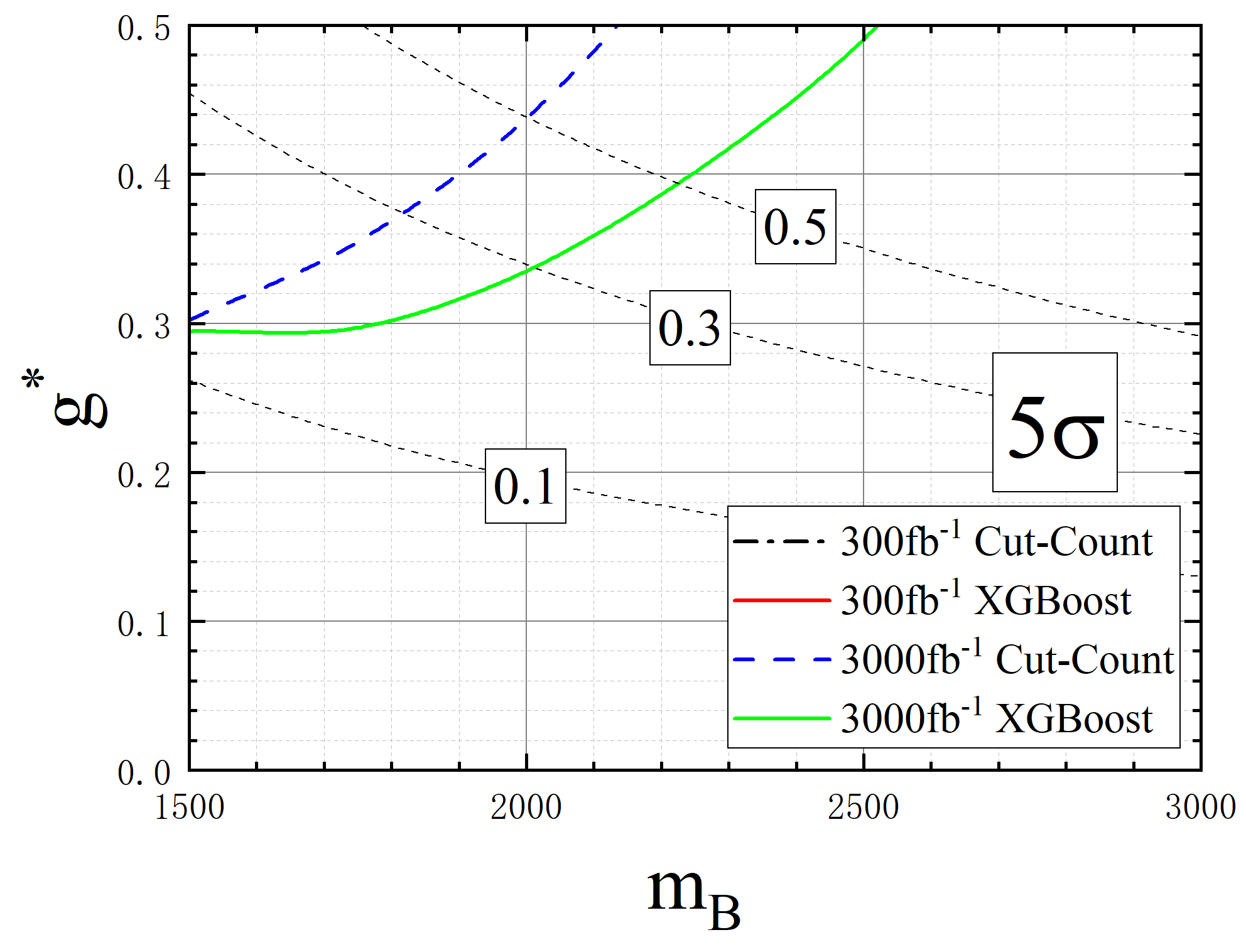}
    \caption{Exclusion (left) and discovery (right) potential for the VLB signal on the $g^{\ast}-m_B$ planes for $R_L=0$ at the 14 TeV LHC, considering the two methods adopted in our analysis. Three typical luminosities, 300 fb$^{-1}$, 1000 fb$^{-1}$ and 3000 fb$^{-1}$, are illustrated. The dashed lines denote the width-to-mass ratios $\Gamma_B/m_B$.}
    \label{RL0}
\end{figure}

In Fig.~\ref{RL0}, we show the $2\sigma$ (exclusion) and $5\sigma$ (discovery) lines on the $g^{\ast}-m_{B}$ plane for $R_L=0$ at the 14 TeV LHC. We can see that  VLBs can be excluded in the regions of 
$g^{\ast}\in \left[0.35,0.50\right]$ and $m_B\in\left[1500 \gev,2300 \gev\right]$ assuming an integrated luminosity of 300 fb$^{-1}$ (Run 3). With a value for the latter of 3000 fb$^{-1}$ (HL-LHC), instead, the exclusion regions can be extended to $g^{\ast}\in \left[0.18,0.50\right]$ 
and $m_B\in\left[1500 \gev,3000 \gev\right]$. Accordingly, the discovery regions are $g^{\ast}\in \left[0.30,0.50\right]$ and $m_B\in\left[1500 \gev,2500 \gev\right]$.
        
\begin{figure}[h!]
        \includegraphics[scale=0.85]{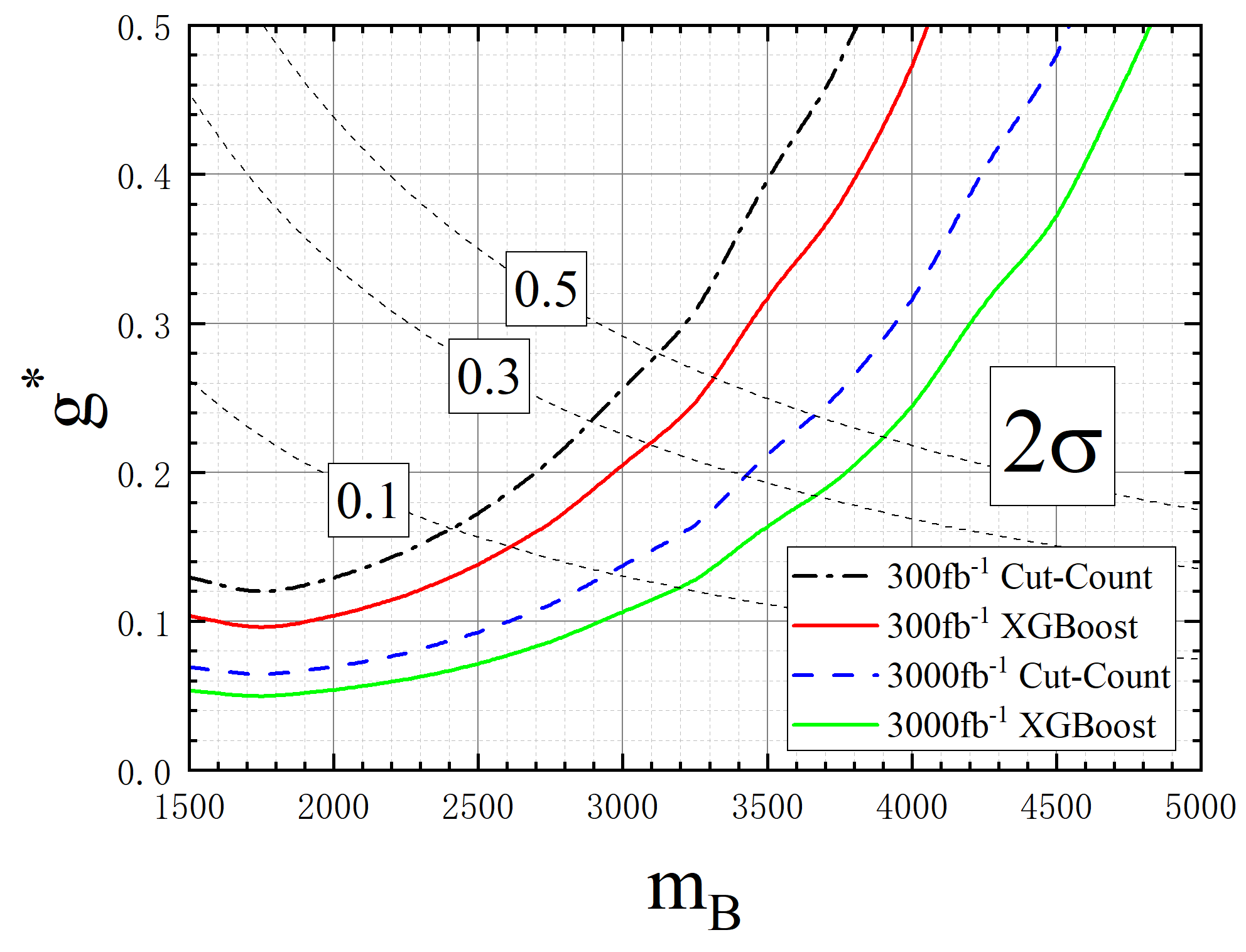}
        \includegraphics[scale=0.85]{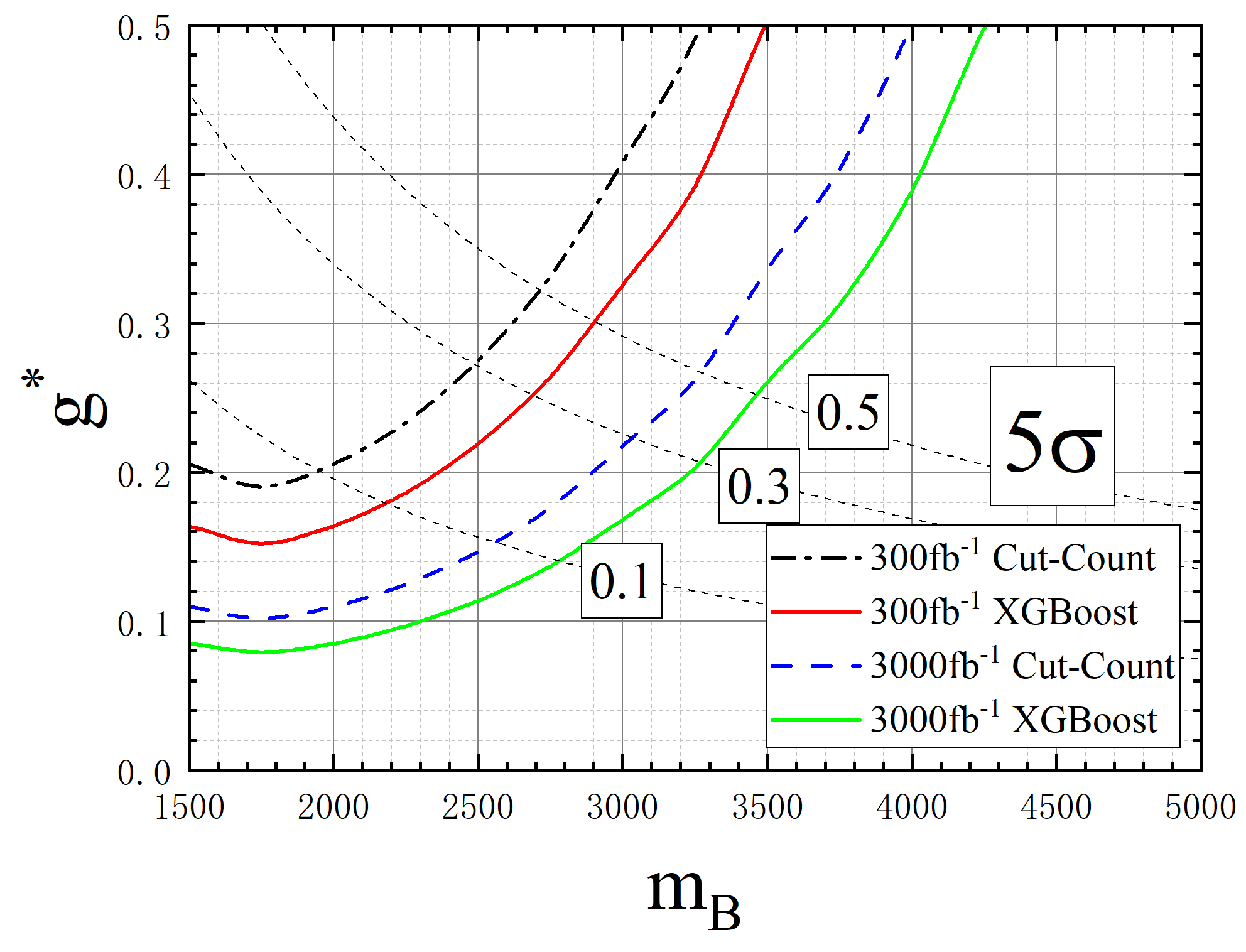}
    \caption{Same as Fig. 6, but for $R_L=0.5$.}\label{RL05}
\end{figure}

Fig.~\ref{RL05} is the same as Fig.~\ref{RL0}, but for $R_L=0.5$. We can see that the exclusion regions are $g^{\ast}\in \left[0.10,0.50\right]$ and $m_B\in\left[1500 \gev,4050 \gev\right]$ with 300 fb$^{-1}$, which can be extended to $g^{\ast}\in \left[0.05,0.50\right]$ and $m_B\in\left[1500 \gev,4750 \gev\right]$ with 3000 fb$^{-1}$. The discovery regions are 
$g^{\ast}\in \left[0.16,0.50\right]$ 
and $m_B\in\left[1500 \gev,3500 \gev\right]$ with 300 fb$^{-1}$ plus  $g^{\ast}\in \left[0.08,0.50\right]$ 
and $m_B\in\left[1500 \gev,4250 \gev\right]$ with 3000 fb$^{-1}$.
        
\section{Summary}

In this work, we have utilized a simplified model approach to test the exclusion and discovery potential of a VLB signal at both the current LHC (Run 3) and a future HL-LHC. We have employed both a standard cut-and-count method and a ML based XGBoost algorithm to achieve this goal. Specifically, we have focused on the process $pp \to B(\to bZ)j\to b(Z \to \nu_l\bar{\nu_l})j$ as signal.

Initially, we have generated signal and background events using state-of-the-art computing tools. Subsequently, we have applied the cut-and-count method and the XGBoost algorithm to classify these events, effectively distinguishing between signal and background. For the XGBoost approach, we have trained the ML model using MC data and utilized Bayesian optimization to determine the optimal combination of hyperparameters. We have then employed this trained model to classify events and determine cut efficiencies via selection thresholds.
Finally, we have evaluated the exclusion and discovery capabilities of the signal over both reducible and irreducible backgrounds across different LHC luminosities (and fixed $\sqrt s=14$ TeV).

Our analysis has revealed significant sensitivities to this VLB process across various LHC stages, spanning from current to future machine configurations. 
These appear to improve significantly upon more traditional  methods used in phenomenological analyses.
\begin{table*}[!t]
    \centering
    \renewcommand{\tablename}{Table}
               	\renewcommand{\arraystretch}{1.5}
               	\vspace{0.4em} \centering
               	\begin{tabular}{ccccccccc}
               		\hline\hline
               		LHC&Decay&\multicolumn{2}{c}{Exclusion} &\multicolumn{2}{c}{Discovery}&Reference \\
               		& & $g^{\ast}$ &$m_{B}$ (GeV) & $g^{\ast}$ &$m_{B}$ (GeV)\\
               		\hline
               		$R_{L}=0$ & $B\to tW$&[0.2,0.5] &[1800,2400] &[0.3,0.5] &[1750,2100] &\cite{Han:2022jcp}  \\
               		$R_{L}=0$ & $B\to bZ$&[0.2,0.5] &[1300,1800] &[0.3,0.5]&[1300,1500]&\cite{Han:2022iqh} \\\hline
               		$R_{L}=0$ & $B\to bZ$&[0.2,0.5] &[1500,3000] &[0.3,0.5] &[1500,2500] &\multirow{2}{*}{this work}  \\
               		$R_{L}$=0.5 &$B\to bZ$&[0.2,0.5] &[3750,4750] &[0.3,0.5] &[3700,4250]  \\
               		\hline
\end{tabular}
\caption{Comparison of exclusion and discovery capabilities of different VLB signals with $\sqrt{s}=14$ TeV and $\mathcal{L}=3000$ fb$^{-1}$ from different works.} \label{other works}
\vspace*{0.5truecm}
\end{table*}
Indeed, for comparison, we finish by listing previous results on searching for a VLB at the HL-LHC in Tab. \ref{other works}, also including $B\to tW$ decays, against ours. Since the advantages of XGBoost method lie in the high-mass region, we focus for this comparison over the region $m_B>$1500 GeV. (Recall that the VLB coupling strength $g^{\ast}$ is also described as $\kappa_{B}$ in other references and their relation is $g^{\ast}=\kappa_{B}$~\cite{Buchkremer:2013bha}.) By comparing these results, we can see that the XGBoost algorithm does improve significantly the sensitivity of searches for a VLB with the same luminosity, a result which makes the exploration of VLB signals ever more promising also from the experimental side. 
However, it should be noted that, due to the inherent characteristics of the ML algorithm, certain limits to its applicability exist. For instance, when the mass of the VLB greatly exceeds the upper bound of the training set (in this work, 4000 GeV), the predictive capability of the model significantly diminishes. In forthcoming research, our primary focus will be on enhancing the model to address this issue.

\section*{Acknowledgements}
We thank Luca Panizzi for helpful discussions. This work of BY, ZL, XJ and LS is supported by the Natural Science Foundation of Henan Province under Grant No. 232300421217, the National Research Project Cultivation Foundation of Henan Normal University under Grant No. 2021PL10, the China Scholarship Council under Grant No. 202208410277 and also powered by the High Performance Computing Center of Henan Normal University. 
The work of SM is supported in part through the NExT Institute, the Knut and Alice Wallenberg
Foundation under the Grant No. KAW 2017.0100 (SHIFT) and the STFC Consolidated Grant No. ST/L000296/1.

\clearpage
\bibliographystyle{utphys}
\bibliography{Refers.bib}

\end{document}